# On the sharing signals measured with DSSD detector

Yu.S.Tsyganov [a], D.Ibadullayev [a,b,c]

[a] FLNR, JINR, 141980 Dubna, Moscow region, Joliot-Curie str.6, Russian Federation

[b] Institute of Nuclear Physics, 050032 Almaty, Kazakhstan

[c] L.N. Gumilyov Eurasian National University, 010000 Astana, Kazakhstan.

tyra@jinr.ru

## Abstract

*In this paper we report on the reasonable nature of sharing signals measured with large area DSSD detectors. This detector manufactured by Micron Semiconductor (UK) and in use at the DGFRS-2 setup installed at DC-280 ultra intense cyclotron of FLNR, JINR. Some results from other setups are presented. We believe that pitch value as well as threshold value play dominant role in the charge sharing process. Some other parameters influenced by that process are considered in brief too, but sometimes in semi-quantity form.*

1. ## Introduction

It was the **DGFRS-2** setup has been put into operation in the end of 2019[th] which allows to use high projectile intensities in the experiments aimed to the synthesis of superheavy elements (SHE) [1-5]. Note that during about last twenty years, new **Z=114-118** SHE were discovered using the **DGFRS** (the **D**ubna **G**as-**F**illed **R**ecoil **S**eparator) setup installed at **U-400 FLNR** cyclotron [6]. The heaviest known atom **Z=118** ( **Oganesson** ) was synthesized in $^{249}Cf+^{48}Ca \rightarrow Og^*$ complete fusion nuclear reaction. From the other hands, to reach for **Z=119, 120** elements in a nearest future, new advanced detector system is strongly required (see e.g. [7-10]).The basic unit of this system is large area DSSD 48x128 strip detector by **Micron Semiconductor** (UK). Nowadays its size is about of 240x60 mm$^2$. During 2020 to 2023 this detector was applied in the different heavy ion induced complete fusion reactions at the **DGFRS-2** setup. In these experiments heavy ions of $^{48}$Ca and $^{40}$Ar were accelerated by DC-280 cyclotron.

## 2. Method of "Active Correlations"

To provide a deep suppression of background products associated with cyclotron beam we apply *"active correlations"* technique during last twenty years.

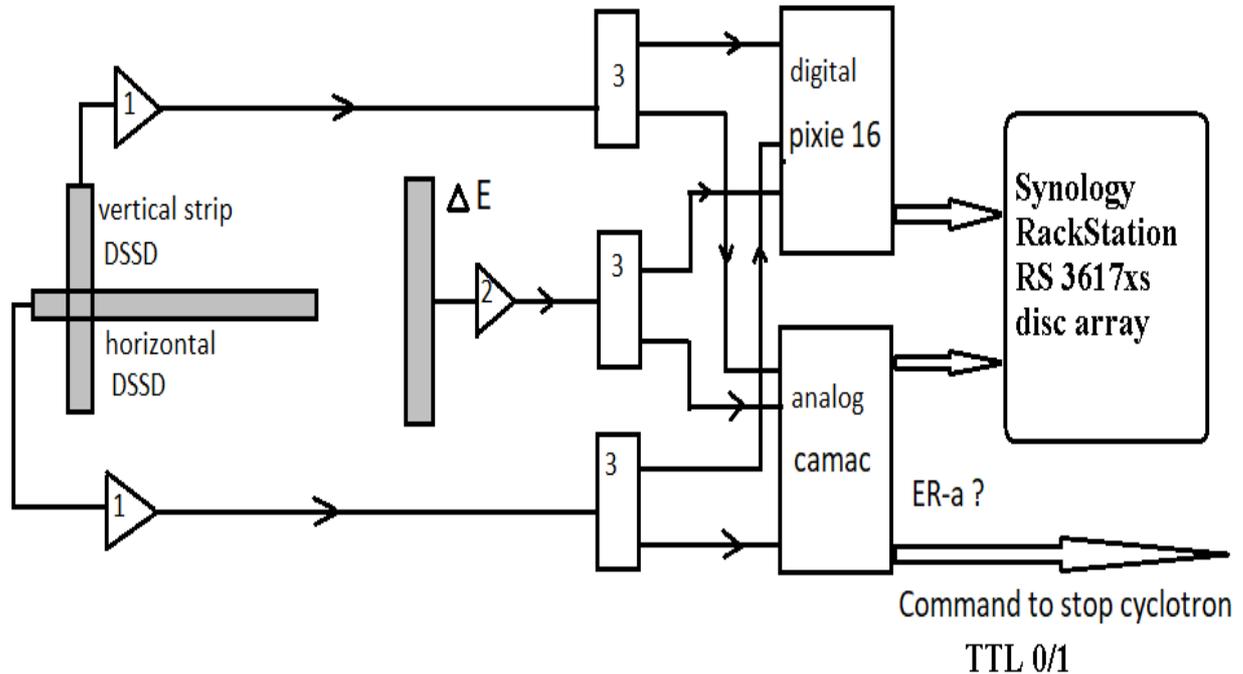

Fig.1 Schematics of the DGFRS-2 spectrometer (1,2-charge sensitive preamplifier, 3- signal splitter)

In the Fig.1 schematics of the **DGFRS-2** two branch spectrometer is shown. Whereas the main data flow is registered by digital branch, namely a **CAMAC** based analog spectrometer provides search for ER-alpha correlation sequences in a real-time mode and generates break point in target irradiation process to provide in fact background free conditions for forthcoming alpha decays registration( see right-down corner on the Fig.1). Note that in the Fig.1 only one vertical and one horizontal strips of **DSSD** detector are shown as well as *ΔE* proportional counter. Both digital and analog data are written to **RS 3617xs** disc array. Additional task for analog spectrometer is to provide on-line monitoring of *ΔE* gaseous pentane filled counter efficiency parameter [11]. The basic object of searching for ER-α sequence in a real-time mode is a "recoil" matrix. Its size is the same as the pixel number of DSSD detector, namely 48x128. As to the element of that matrix it is filled by elapsed time value of the registered recoil. But, if we detected two sharing signals, the same elapsed time parameter is written to two neighbor elements. Sharing signals from front (48) strips are ignored in the sense that those signals not take part in real-time search of ER-alpha sequences. But they are written in file and can be processed in off-line analysis. In contrast to back side signals which definitely take part in that method of "active correlations". In the text it is mentioned when those signals are detected, the elapsed times are written to both cells of ER-matrix located in the PC RAM ( i, j; i, j+1 e.g. where i- index of front strip, j-back one). Note that sharing back signals are writes as one event, where as sharing front signals as individual events following by each other. As example, event with ΔE > 0 and e1 and e2 in front neighbor strips will be as:

- E1, ΔE >0, y1>0, t1 in first event;
- E2, ΔE=0, y1=0; t1+t(dead) in the next event, where t(dead)~25 μs
  Of course, in more sophisticated nearest future algorithm it will be possible in principle to take it into account, but, due to this effect is negligible, this possibility is not in use now.

Here, y1 – back strip number. It should be noted that a present version of **Builder C++ YDA** program [12, 13] provides on-line visualization of sharing signal number (see Fig.2).

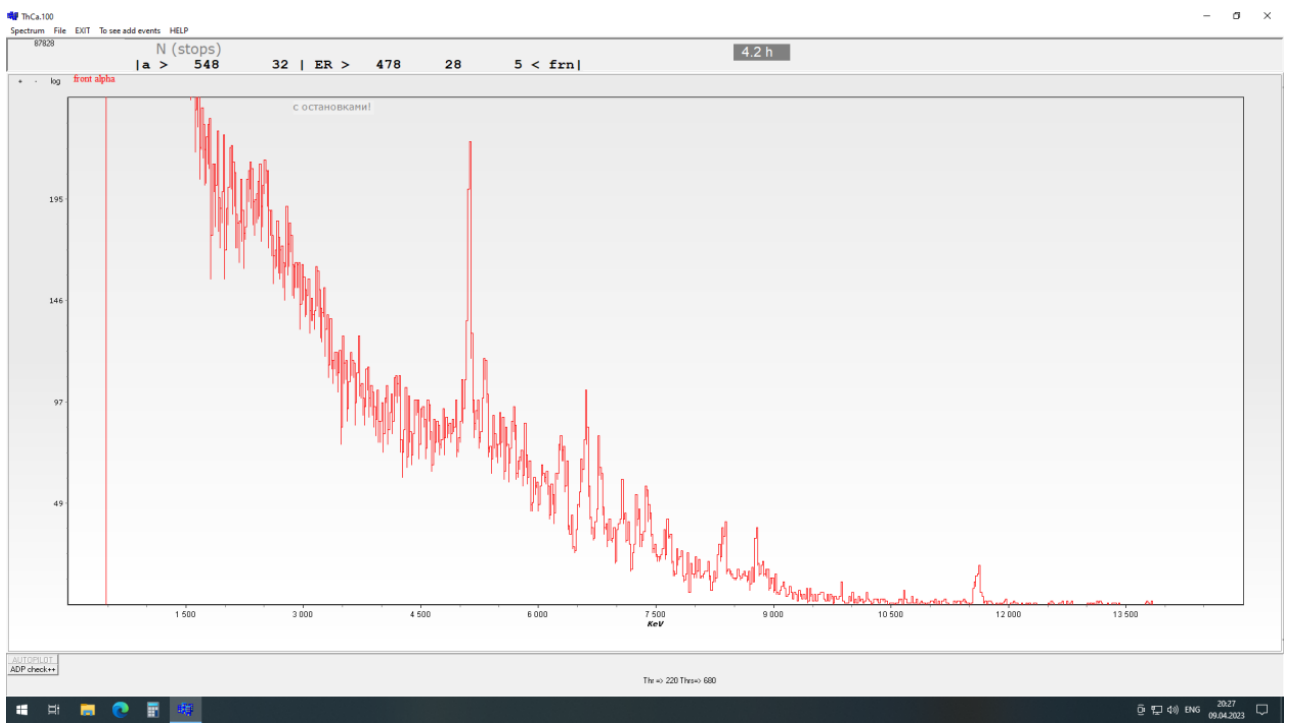

Fig.2 Typical on line α-spectrum picture of YDA C++ program. Numbers of sharing signals are shown in upper line for 4.2 hours of acquisition

In this figure it is shown: 32 sharing signals like "α" (totally 548) and 28 ones like "ER" (totally 478) for back 128 strips, whereas 5 sharing "ER" signals for front 48 strips.

## 3. Sharing signal registering with DSSD detector

In the **Table 1** results of registering of sharing signals are presented. Of course, the condition of these signals formation are multifunctional, but to a first approximation one can state that threshold value and pitch size one can consider like main incoming parameters. In the Fig.3 the dependence of percentage value against the pitch size variable is shown. All detectors were operated in totally depletion mode in heavy ion induced complete fusion nuclear reactions. Bias supply voltage in each case was about 40-60 volts.

**Table 1. Sharing signal measured parameters for back strips**

| variable | Present detector DGFRS-2 | BB17 DGFRS-2 | [16] | DGFRS-3 setup, ттт-20Micron. Scd. (cooled detector) [14] | Present digital sys. | TASCA, GSI [15] |
|---|---|---|---|---|---|---|
| pitch | 2 | 1 | 0.76 | 0.76 | 2 | 1 |
| alfa 9160-9360 | 5.4% (13486; 732) | ~8% | 17% | ~100% 27% (front strips) | 7.2% | |
| ER 8000 – 18000 | 7.4% (485447; 36052) | | | | 9.1 % | ~15% all |
| Threshold, KeV | 700-800 | 700-800 | 700 | 150/40 | ~300 | 300-500 |
| Recalculated | 6.4% | 8*6.4/5.4= 9.4% | 17%*700/750 = 15.87 % | 150/750*100= 20% | 9.1*300/750= 3.6 % | 15*400/700 =8.6% |
| Pitch*<prc> | **12.8** | **9.4** | **12.1** | **15.2** | **7.2** | **8.6** |

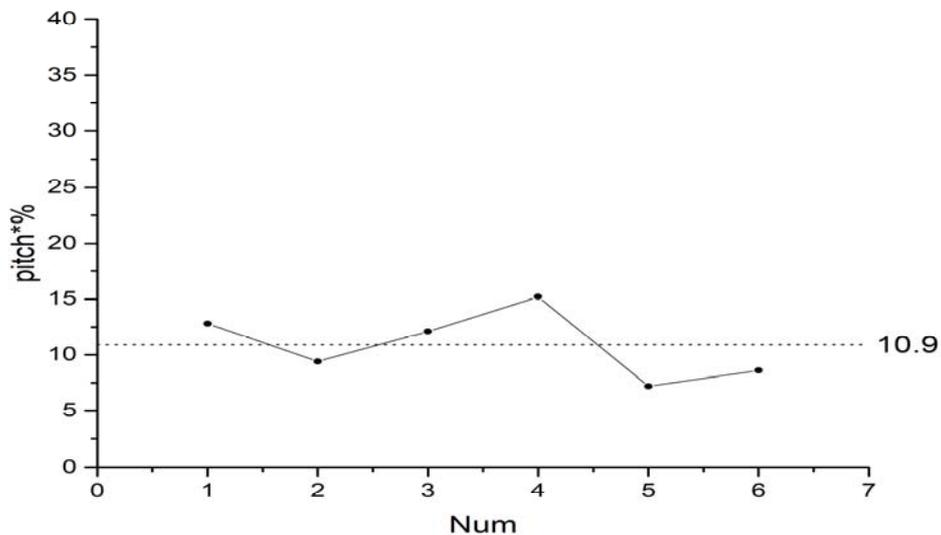

Fig.3 Dependence of pitch width times percentage against column number <mean>=10.9% STD=2.99%. (Back strips)

In the Fig.4 the bias voltage dependence is shown for ~5.5 MeV alpha particles.

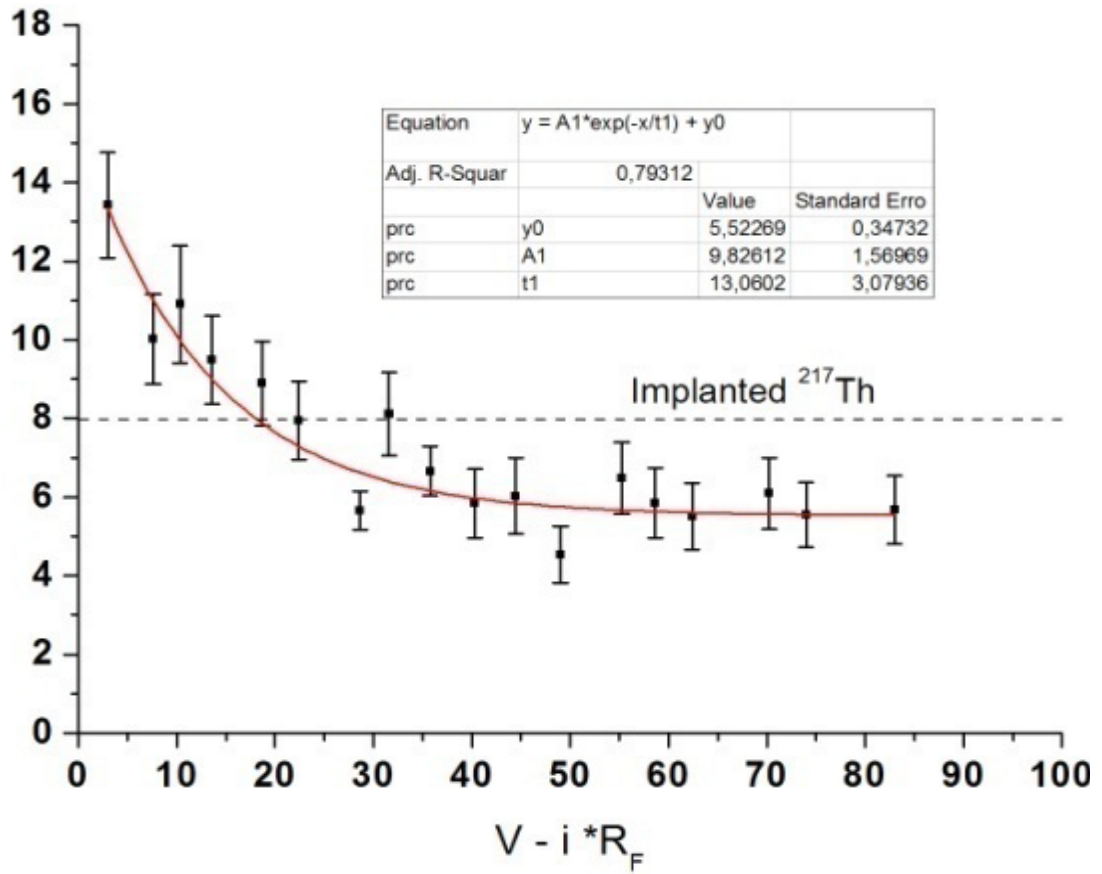

Fig.4 Percentage ( %) for alpha ~5.5 MeV (ext. source) back strips DSSD thr~800 KeV. Implanted $^{217}$Th line (dashed) corresponds to $^{nat}Yb+^{48}Ca \rightarrow ^{217}Th+3n$ complete fusion nuclear reaction

## 4. Summary

Pitch size as well as threshold value play dominant role in the formation of sharing signals measured with DSSD detector.

From the other side (as far as we can state) more number questions than answers according to the effect. Reasonably, higher statistics of applications of different (pitch, depth, voltage, A-Z, range, etc…) detectors are required.

## 5. Supplement1. A role of charge particle range(semi-qualitevely)

Below, on Fig.5a,b and using Einstein relationship mobility-diffusion coefficient in silicon, one can see some difference in formation of sharing signals for relatively short and long path particles. This approach explains qualitively difference in the effect for recoils (relatively short, units of microns) and α-particles(long, tens of microns).

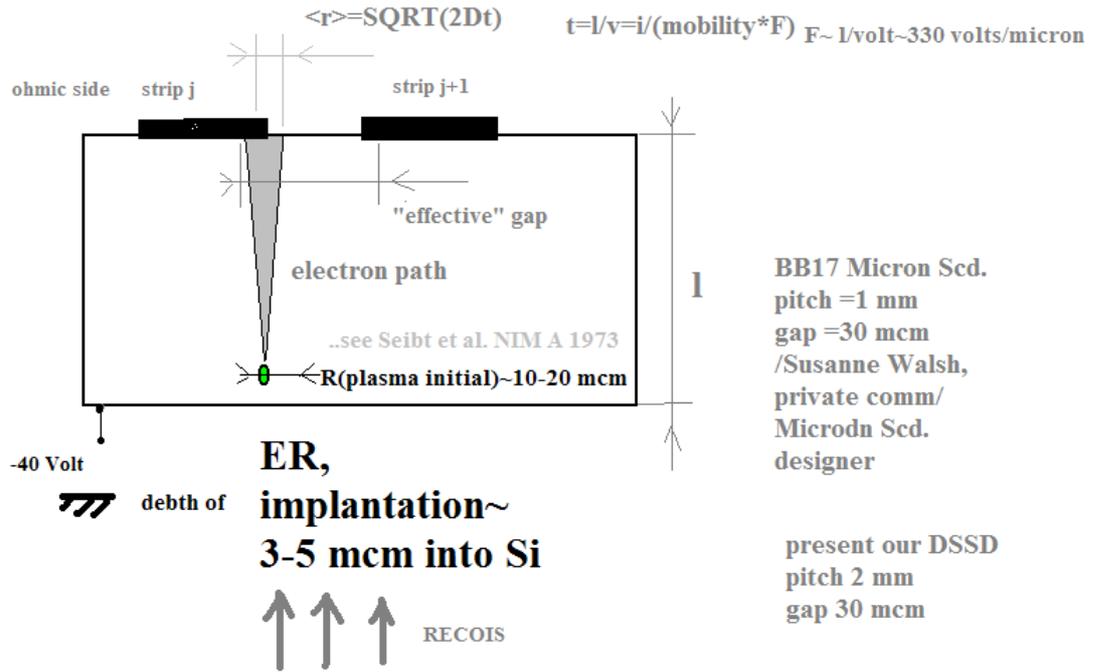

Fig.5a  A broadening of electron path size due to diffusion(random motion..) to explain larger effective gap area

Here, diffusion coefficient for electrons related with its mobility by Einstein relationship:

$D_e = \mu_e k_b T$, where $\mu_e$ - electron mobility in silicon, $k_b$ - Boltzmann constant.

$\mu_e = (1400...1900)$ cm$^2$/(Volt·s)

Additionally, type (range) of ionizing particle should be also to be as an input parameter, although it is not clear direction of influencing to effect.

For instance, alpha particle 5.5-11 MeV has greater range value than recoil one, so the "centroid" of plasma column is more close in average to back strip contact(Fig.5b).

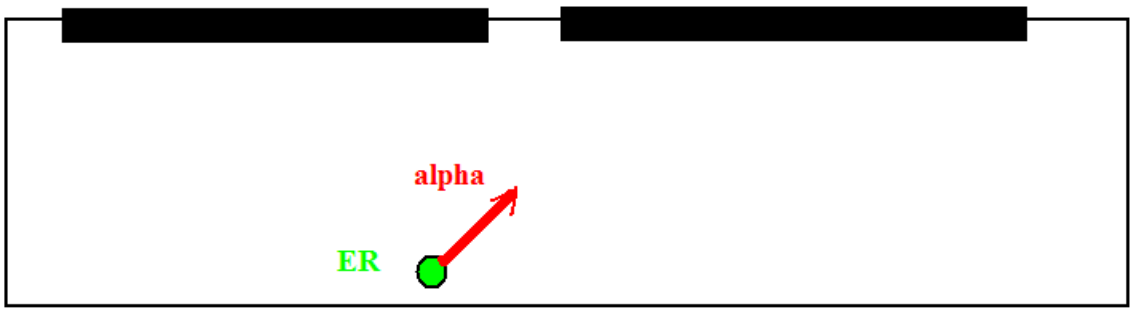

Fig.5b  Schematics for both short and long path particles detection with DSSD of ~300 µm Si substrate thickness (neighbor back strips-upper part is shown in black)

## 6. Supplement 2 Resistive layer sharing model

In Ref.'s [9, 10] back strip sharing signals are obey to the equation of:

$E_1 + E_2 = const = E_0 - \Delta$.

Here: $E_{1,2}$ are sharing signals, $E_0$ is corresponded a single strip signal, and $\Delta$-small ballistic deficit of about ~29 KeV (for $^{217}$Th isotope). Note, that in [17] authors considered namely resistive like signal division model too.